\documentclass[twocolumn,american]{revtex4-2}
\usepackage[latin9]{inputenc}
\usepackage{babel}
\usepackage{amsmath}
\usepackage{amssymb}
\usepackage{graphicx}
\usepackage[pdfusetitle,
 bookmarks=true,bookmarksnumbered=false,bookmarksopen=false,
 breaklinks=false,pdfborder={0 0 1},backref=false,colorlinks=false,pdfpagemode=FullScreen]
 {hyperref}
\usepackage{lineno}
\linenumbers

\makeatletter

\newcommand{\lyxmathsym}[1]{\ifmmode\begingroup\def\b@ld{bold}
  \text{\ifx\math@version\b@ld\bfseries\fi#1}\endgroup\else#1\fi}

\usepackage{dcolumn}
\usepackage{bm}
\usepackage{wrapfig}
\usepackage{subfigure}
\usepackage{ucs}
\usepackage{epsfig}
\usepackage{psfrag}\usepackage{epstopdf}
\DeclareGraphicsExtensions{.pdf,.eps,.png,.jpg,.mps}

\usepackage{babel}

\nolinenumbers 

\makeatother

\begin{document}
\title{Thermodynamic and magnetocaloric properties of a triangular spin-1/2
cluster with Dzyaloshinskii-Moriya interaction}
\author{Jordana Torrico$^{1}$, Rômulo A. Silva$^{2}$, Sergio M. de Souza$^{2}$,
Onofre Rojas$^{2}$}
\affiliation{$^{1}$ Instituto de Ciências Exatas, Universidade Federal de Alfenas,
37133-840 Alfenas, MG, Brazil}
\affiliation{$^{2}$ Department of Physics, Institute of Natural Science, Federal
University of Lavras, 37200-900 Lavras-MG, Brazil}
\begin{abstract}
We present a theoretical investigation of the magnetic and thermodynamic
properties of the triangular spin-1/2 cluster with Dzyaloshinskii-Moriya
(DM) interaction, described by a spin-1/2 Heisenberg Hamiltonian with
antisymmetric exchange interactions. The energy spectrum and ground-state
phase diagram reveal the presence of ferromagnetic (FM), ferrimagnetic
(FI), and frustrated (FR) phases, strongly influenced by the total
spin and the DM interaction. We analyze magnetization and susceptibility,
showing that at low temperatures the system exhibits a characteristic
1/3 magnetization plateau, while thermal fluctuations suppress magnetic
order at higher temperatures. The entropy and specific heat display
residual entropies due to ground-state degeneracies, Schottky-type
anomalies at intermediate temperatures, and additional low-temperature
features related to phase transitions. Particular attention is given
to the magnetocaloric effect (MCE), characterized by both direct and
inverse regimes depending on the magnetic field variation. We find
that the DM interaction enhances the complexity of the MCE, leading
to nontrivial entropy variations as a function of the magnetic field.
These results provide insights into the role of frustration and anisotropy
in tuning the MCE of properties triangular spin clusters, with relevance
to {\normalsize$\mathrm{Cu}_{3}$}-based molecular magnets.
\end{abstract}
\maketitle

\section{Introduction}\label{intro}

Molecular nanomagnetism has attracted significant attention over the
past decades \citep{gatteschi,sessoli}. This field encompasses the
synthesis, characterization, and theoretical modeling of molecular
magnetic materials with unique properties and diverse dimensionalities.
Molecular magnets, defined by their well-localized magnetic moments,
provide a versatile platform for exploring fundamental aspects of
quantum mechanics and for testing theoretical models \citep{maria}.
Beyond their fundamental interest, these systems are also promising
candidates for nanoscale technological applications \citep{leuenberger,lehmann,thomas}.
Their magnetic characteristics and tunability enable potential uses
across physics, magnetochemistry, biology, biomedicine, and materials
science \citep{ses,gat,coro}. In addition, molecular magnets provide
a platform for exploring a variety of phenomena, including quantum
tunneling of magnetization\citep{thomas,friedman,julien}, quantum
computing and spintronics\citep{leuenberger,carretta,bogani,clemente,cororev,wang,wang2},
molecular-scale magnetic memory\citep{ses}, quantum interference
\citep{wern}, and the occurrence of level crossings and magnetization
plateaus \citep{julien,taft}.

Among the potential applications, the magnetocaloric effect (MCE)
is particularly noteworthy. The MCE refers to the heating or cooling
of a magnetic material under the application of an external magnetic
field, a property that can be exploited in the development of environmentally
friendly refrigeration technologies \citep{franco,rome}. A wide range
of magnetic systems, including molecular magnets \citep{rome,tishin,karol,karol1,torrico,torrico1},
exhibit a significant magnetocaloric response, making them highly
attractive for next-generation magnetic refrigerators. The search
for materials with enhanced MCE properties has therefore highlighted
molecular magnets as especially promising candidates for nanoscale
refrigeration \citep{rsessoli,zheng}. Fundamentally, the MCE originates
from the dependence of the magnetic entropy on both temperature and
the applied field \citep{franco}. In particular, geometrically frustrated
systems \citep{schnack,schnack1} display pronounced MCE near field-induced
transitions, especially of first-order type. In the conventional MCE,
the magnetic entropy decreases as the external field is increased
isothermally. Conversely, in certain frustrated ferrimagnetic systems,
an inverse magnetocaloric effect has been reported: under adiabatic
magnetization, the entropy increases and the material undergoes cooling
\citep{Ranke,Ranke1,Florez,Szalowski}.

The MCE has been reported in several molecular magnets. For $\mathrm{V}_{6}$\},
exact numerical studies within the Heisenberg model revealed a highly
tunable isothermal entropy change, including regions of inverse MCE
\cite{karol1,torrico}. The molecule $\mathrm{V}_{12}$ also exhibits
both conventional and inverse MCE, with Szalowski and Kowalewska \cite{karol}
showing that quantum level crossings produce alternating regimes,
the strongest inverse response occurring at cryogenic temperatures.
In $\mathrm{Cu}_{5}$-NIPA, Szalowski \cite{karol2} demonstrated
a strong field dependence of entropy and specific heat and a wide
tunability of the isothermal entropy change, while Torrico and Plascak
\cite{torrico1} confirmed the presence of inverse MCE. More recently,
triangular Cu-based clusters have attracted growing attention: Rojas
et al. studied magnetization plateaus and entanglement in Cu(II) complexes
\cite{Mirzoyan}. More recently, quantum machines based on a $\mathrm{Cu}_{3}$-like
Heisenberg system were proposed \cite{moises}, and Antonio et al.
analyzed thermodynamic responses and the MCE in $\mathrm{Cu}_{3}$
triangular models with Dzyaloshinskii--Moriya (DM) interactions \cite{gilberto}. 

The triangular geometry provides a particularly rich framework for
investigating the magnetic properties of molecular magnets. Systems
based on $\mathrm{Cu}^{2+}$ ions, which carry spin-1/2, have been
extensively studied experimentally \citep{lida,spielberg,kintzelg,liu}.
Spin interactions on triangular units can give rise to both frustrated
and non-frustrated configurations, which explains why triangular motifs
are among the most widely examined in molecular magnetism. Examples
include $\mathrm{V}_{3}$ \citep{ponomaryov,choi,choi1,luzon,kortz},
$\mathrm{V}_{5}$ \citep{karol,nath}, $\mathrm{V}_{6}$ \citep{karol1,torrico,luban,haraldsen},
and $\mathrm{V}_{15}$ \citep{fu,kosty,konstanti}. The Heisenberg
spin triangle serves as a powerful minimal model for capturing the
mechanisms of single-molecule magnetism, including the case of $\mathrm{Cu}_{3}\mathrm{(OH)}$\citep{ponomaryov,choi,choi1,luzon,kortz}.
Owing to its structural simplicity combined with inherent frustration,
the triangular model offers a promising platform for exploring the
magnetocaloric effect and other unconventional magnetic and thermodynamic
phenomena. The role of DM interactions in triangular molecular clusters
has been extensively discussed in the literature, particularly in
systems such as $\mathrm{V}_{15}$, where anisotropic interactions
play a crucial role in determining the low-energy spectrum and magnetic
anisotropy\citep{choi,fu,kosty,konstanti} .

Inelastic neutron scattering (INS) experiments on powder samples of
the triangular molecular nanomagnet $\mathrm{Cu}_{3}$ were reported
by Iida, Qiu, and Sato \citep{lida}. These measurements revealed
two excitations at 0.5 and 0.6 meV, along with an additional low-energy
peak at 0.1 meV. Based on these observations, a Hamiltonian model
was proposed and optimal exchange parameters were determined. The
presence of the 0.1 meV mode provided direct evidence of DM interactions
in $\mathrm{Cu}_{3}$. Moreover, the spin-lattice coupling was found
to be exceptionally weak, leading to either rigid spin states or long
spin lifetimes at low temperatures. Remarkably, the inelastic peaks
persisted up to very high temperatures, further demonstrating that
the coupling between spin and lattice vibrations in $\mathrm{Cu}_{3}$
is much weaker than in other known molecular nanomagnets.

Therefore, in this work we investigate the magnetic and thermodynamic
properties of a spin-1/2 Heisenberg triangle, which serves as a minimal
model for triangular Cu-based molecular clusters \citep{ponomaryov,choi,choi1,luzon,kortz}.
The paper is organized as follows. In Sec. \ref{model}, we introduce
the Hamiltonian for the triangular spin-1/2 cluster, present the exact
analytical solution of its energy spectrum and eigenstates, and discuss
the corresponding ground-state phase diagram. In Sec. \ref{sec:Res},
we analyze the magnetic and thermodynamic behavior of the system,
focusing on magnetization, susceptibility, entropy, specific heat,
and the magnetocaloric effect. Finally, Sec. \ref{sec:Clsn} summarizes
our main findings and provides concluding remarks.

\section{Model and phase diagram}\label{model}

\begin{figure}[h]
\includegraphics[scale=0.35]{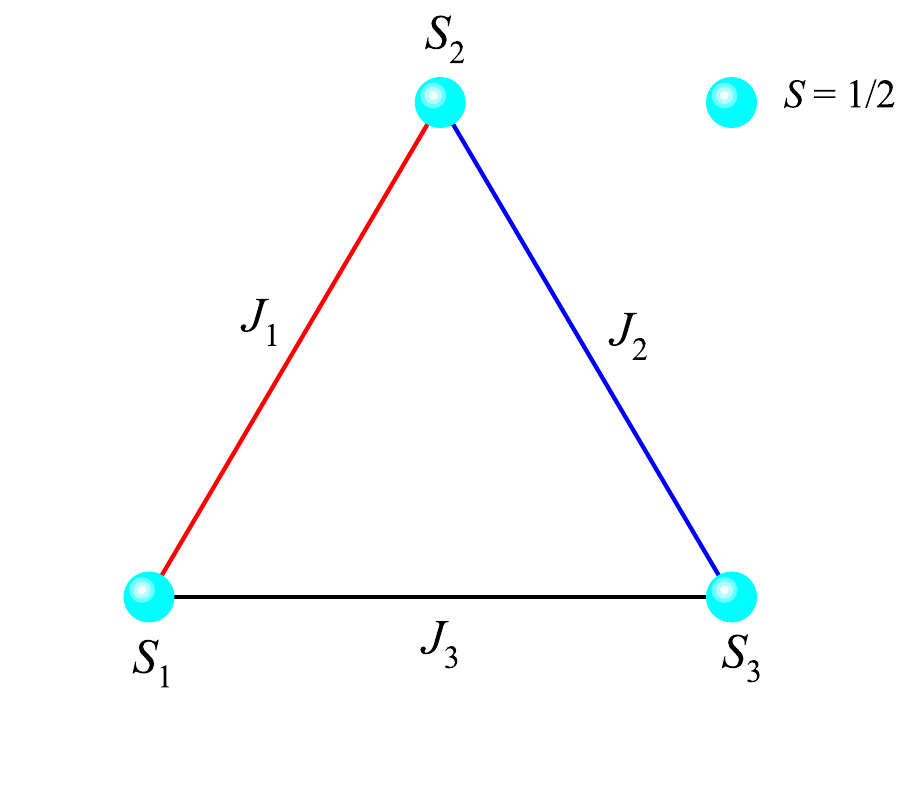} \caption{ Schematic representation of a spin-1/2 Heisenberg triangle. The
magnetic ions (representing $\mathrm{Cu}^{2+}$-like spins) are shown
as light blue circles, and the Heisenberg exchange interactions between
neighboring spins are denoted by $J_{1}$, $J_{2}$, and $J_{3}$.}\label{fig:1}
\end{figure}

We consider a spin-1/2 Heisenberg triangle motivated by $\mathrm{Cu}^{2+}$-based
molecular clusters, where three magnetic ions form a triangular arrangement
with Cu$^{2+}$ ions located at its vertices, as schematically illustrated
in Fig.\ref{fig:1}. Its magnetic behavior can be described by a spin-1/2
Heisenberg triangle, governed by the Hamiltonian
\begin{align}
\mathcal{\hat{H}}=\! & -\!\sum^{3}_{j=1}\!\left[J_{j}\vec{S}_{j}\cdot\vec{S}_{j+1}\!+\!BS^{z}_{j}\!+\!\vec{D}_{j}\!\cdot\!(\vec{S}_{j}\times\vec{S}_{j+1})\right]\!,\label{ham}
\end{align}
where $\vec{S}=(S^{x}_{j},S^{y}_{j},S^{z}_{j})$ denote the spin-1/2
operator at site $j$. Here and in the following, periodic boundary
conditions are assumed, such that $\vec{S}_{4}\equiv\vec{S}_{1}$.
The exchange couplings $J_{1}$, $J_{2}$, $J_{3}$ describe interactions
along the three edges of the triangle: between sites (1-2), (2-3),
and (3-1), respectively. The external magnetic field applied along
the $z$-axis is given by $B=g\mu_{B}h$, where $g$ is the Landé
g-factor and $\mu_{B}$ is the Bohr magneton. The Landé g-factor is
a dimensionless parameter characterizing the magnetic response of
the $\mathrm{Cu}^{2+}$ ion, while the Bohr magneton $\mu_{B}$ is
the fundamental unit of electronic magnetic moment. In general, the
exchange couplings need not be identical: while the symmetric case
$J_{1}=J_{2}=J_{3}$ corresponds to an ideal equilateral triangle,
real molecular clusters often exhibit distortions due to ligand environments,
leading to inequivalent exchange paths\citep{spielberg,kintzelg,liu}.
Therefore, considering $J_{1}\neq J_{2}\neq J_{3}$ allows us to describe
more realistic situations and to investigate the effect of symmetry
breaking on the magnetic and thermodynamic properties of triangular
clusters \citep{ponomaryov,choi,choi1,luzon,kortz}.

Finally, the DM vectors $\vec{D}_{j}$ account for antisymmetric exchange
interactions proportional to the vector product of neighboring spins.
This interaction originates from spin-orbit coupling in systems lacking
inversion symmetry at the bond center, as established by Moriya\textquoteright s
theory. In triangular $\mathrm{Cu}^{2+}$-based clusters, the combination
of spin-orbit effects and low-symmetry ligand environments can give
rise to finite DM interactions, which have been experimentally observed
in triangular molecular clusters \citep{lida,choi,choi1,luzon,kortz}.

In general, symmetry considerations constrain the orientation of the
DM vectors; for triangular clusters, they are often oriented perpendicular
to the molecular plane, $\vec{D}_{j}=(0,0,D_{z})$. However, in order
to obtain closed analytical expressions and to capture the overall
effect of antisymmetric exchange in a simplified manner, we adopt
an effective isotropic parametrization $D_{x}=D_{y}=D_{z}=D$. This
choice should be understood as a minimal model that preserves the
qualitative impact of the DM interaction, rather than a fully symmetry-constrained
microscopic description.

For the symmetric case $J_{1}=J_{2}=J_{3}$ and $D=0$, the Hamiltonian
is invariant under the permutation group $S_{3}$, equivalently associated
with the trigonal symmetry of the triangular cluster. In this situation,
the three-spin Hilbert space decomposes into one quartet with total
spin $S=3/2$ and two degenerate doublets with total spin $S=1/2$.
This degeneracy is not accidental, but follows directly from the trigonal
symmetry of the molecule\citep{schnack,schnack1,konstanti}. Group-theoretical
analyses of triangular exchange multiplets {[}see, e.g., Refs. \citep{lida,choi,choi1,luzon,kortz}
and related reviews{]} show that these two lowest-energy doublets
form a doubly degenerate $E$ representation of the trigonal group.
The corresponding states are associated with opposite chiralities
of the spin circulation around the triangle and may therefore be interpreted
as an effective orbital degree of freedom with an unquenched orbital
moment.

In the presence of spin-orbit coupling, this orbital sector naturally
couples to the spin degrees of freedom, generating magnetic anisotropy
and antisymmetric exchange interactions of Dzyaloshinskii-Moriya type.
Consequently, the DM interaction provides a natural mechanism for
lifting the chiral doublet degeneracy, inducing spin chirality, stabilizing
non-collinear magnetic configurations, and modifying the low-energy
spectrum of triangular molecular magnets. In the present work, we
focus on the antisymmetric exchange as the leading source of magnetic
anisotropy. Other anisotropic contributions, such as symmetric exchange
anisotropy, are neglected for simplicity, while single-ion anisotropy
is absent for spin-1/2 systems.

By diagonalizing the Hamiltonian \eqref{ham}, the following eigenvalues
are obtained: 
\begin{align}
\varepsilon_{1}= & -\frac{J_{1}}{4}-\frac{J_{2}}{4}-\frac{J_{3}}{4}-\frac{3B}{2},\label{eq:e1}\\
\varepsilon_{2}= & -\frac{J_{1}}{4}-\frac{J_{2}}{4}-\frac{J_{3}}{4}+\frac{3B}{2},\\
\varepsilon_{3}= & -\frac{J_{1}}{4}-\frac{J_{2}}{4}-\frac{J_{3}}{4}-\frac{B}{2},\label{eq:e3}\\
\varepsilon_{4}= & -\frac{J_{1}}{4}-\frac{J_{2}}{4}-\frac{J_{3}}{4}+\frac{B}{2},\\
\varepsilon_{5}= & \frac{J_{1}}{4}+\frac{J_{2}}{4}+\frac{J_{3}}{4}+\frac{1}{2}K_{+},\\
\varepsilon_{6}= & \frac{J_{1}}{4}+\frac{J_{2}}{4}+\frac{J_{3}}{4}-\frac{1}{2}K_{+},\label{eq:e6}\\
\varepsilon_{7}= & \frac{J_{1}}{4}+\frac{J_{2}}{4}+\frac{J_{3}}{4}+\frac{1}{2}K_{-},\\
\varepsilon_{8}= & \frac{J_{1}}{4}+\frac{J_{2}}{4}+\frac{J_{3}}{4}-\frac{1}{2}K_{-},\label{eq:e8}
\end{align}
with 
\begin{align*}
K_{\pm}= & \sqrt{\left(B\pm\sqrt{3D^{2}+J^{2}_{s}}\right)^{2}+6D^{2}},\\
J^{2}_{s}= & J^{2}_{1}+J^{2}_{2}+J^{2}_{3}-J_{1}J_{2}-J_{1}J_{3}-J_{2}J_{3}.
\end{align*}

To analyze the thermodynamic and magnetic properties of the system,
we evaluate the partition function of the spin-1/2 Heisenberg triangle
using the eigenvalues given in Eqs.(\ref{eq:e1}-\ref{eq:e8}). This
yields

\begin{alignat}{1}
\mathcal{Z}= & 2{\rm e}^{-\frac{J_{1}+J_{2}+J_{3}}{4k_{B}T}}\left[{\rm ch}\left(\tfrac{K_{+}}{2k_{B}T}\right)+{\rm ch}\left(\tfrac{K_{-}}{2k_{B}T}\right)\right]\nonumber \\
 & +2{\rm e}^{\frac{J_{1}+J_{2}+J_{3}}{4k_{B}T}}\left[{\rm ch}\left(\tfrac{3B}{2k_{B}T}\right)+{\rm ch}\left(\tfrac{B}{2k_{B}T}\right)\right],\label{z}
\end{alignat}
where $k_{B}$ is the Boltzmann constant and $T$ is the absolute
temperature. Eq.\eqref{z} corresponds to the canonical partition
function of the model. The Helmholtz free energy then follows as 
\begin{eqnarray}
F(T,B)=-k_{B}T\ln(\mathcal{Z}).
\end{eqnarray}
From the free energy, the main thermodynamic and magnetic quantities
can be derived. In particular, we consider the magnetization $M=-\left(\tfrac{\partial F}{\partial B}\right)_{T}$,
the magnetic susceptibility $\chi=\left(\tfrac{\partial M}{\partial B}\right)_{T}$,
the entropy $\mathcal{S}=-\left(\tfrac{\partial F}{\partial T}\right)_{B}$,
and the specific heat $C=T\left(\tfrac{\partial\mathcal{S}}{\partial T}\right)_{B}$.

The MCE describes the change in temperature that occurs when the external
magnetic field is varied. It can be characterized in two complementary
ways: (i) the adiabatic temperature change, where the system undergoes
a variation in temperature without heat exchange with the environment,
and (ii) the isothermal entropy change, where the temperature is kept
constant while the entropy varies with the applied field.

An alternative way to characterize the magnetocaloric effect is through
the change in magnetic entropy. Magnetic entropy quantifies the degree
of magnetic disorder in the system and is strongly influenced by the
application of an external magnetic field. A variation in the field
modifies the magnetic entropy, which in turn may lead to a temperature
change. Thus, monitoring the entropy variation provides valuable insight
into the MCE.

The isothermal entropy change can be written using a Maxwell relation,
$\left(\frac{\partial M}{\partial T}\right)_{B}=\left(\frac{\partial S}{\partial B}\right)_{T}$,
leading to
\begin{eqnarray}
\Delta\mathcal{S}(T,\Delta B) & = & \int^{B_{f}}_{B_{i}}\left(\dfrac{\partial M(T,B)}{\partial T}\right)_{B}dB\nonumber \\
 & = & \mathcal{S}(T,B_{f})-\mathcal{S}(T,B_{i}),\label{mce}
\end{eqnarray}
where $\Delta B=B_{f}-B_{i}$ is the change in the applied magnetic
field, with $B_{i}$ and $B_{f}$ denoting the initial and final fields,
respectively. According to the literature \citep{franco,pech}, we
adopt the convention that a positive $-\Delta\mathcal{S}>0$ corresponds
to the conventional (direct) MCE, while a negative $-\Delta\mathcal{S}<0$,
indicates the inverse MCE.

We stress that the qualitative effects discussed below, such as the
lifting of degeneracies and the modification of thermodynamic responses,
are generic consequences of the DM interaction and are not restricted
to the isotropic parametrization adopted here.

\section{Results and discussion}\label{sec:Res}

In the following sections, we present and analyze the energy spectrum
and ground-state phase diagram of the model, together with its magnetic
and thermodynamic properties, including magnetization, magnetic susceptibility,
entropy, and specific heat. Particular emphasis is placed on the magnetocaloric
effect. For convenience, we set $k_{B}=1$, so that $k_{B}T\rightarrow T$,
and also take $\mu_{B}=1$ and $g=1$. These conventions will be used
throughout the remainder of this work.

\subsection{Energy spectrum and phase diagram}

\begin{figure}[h]
\includegraphics[scale=0.4]{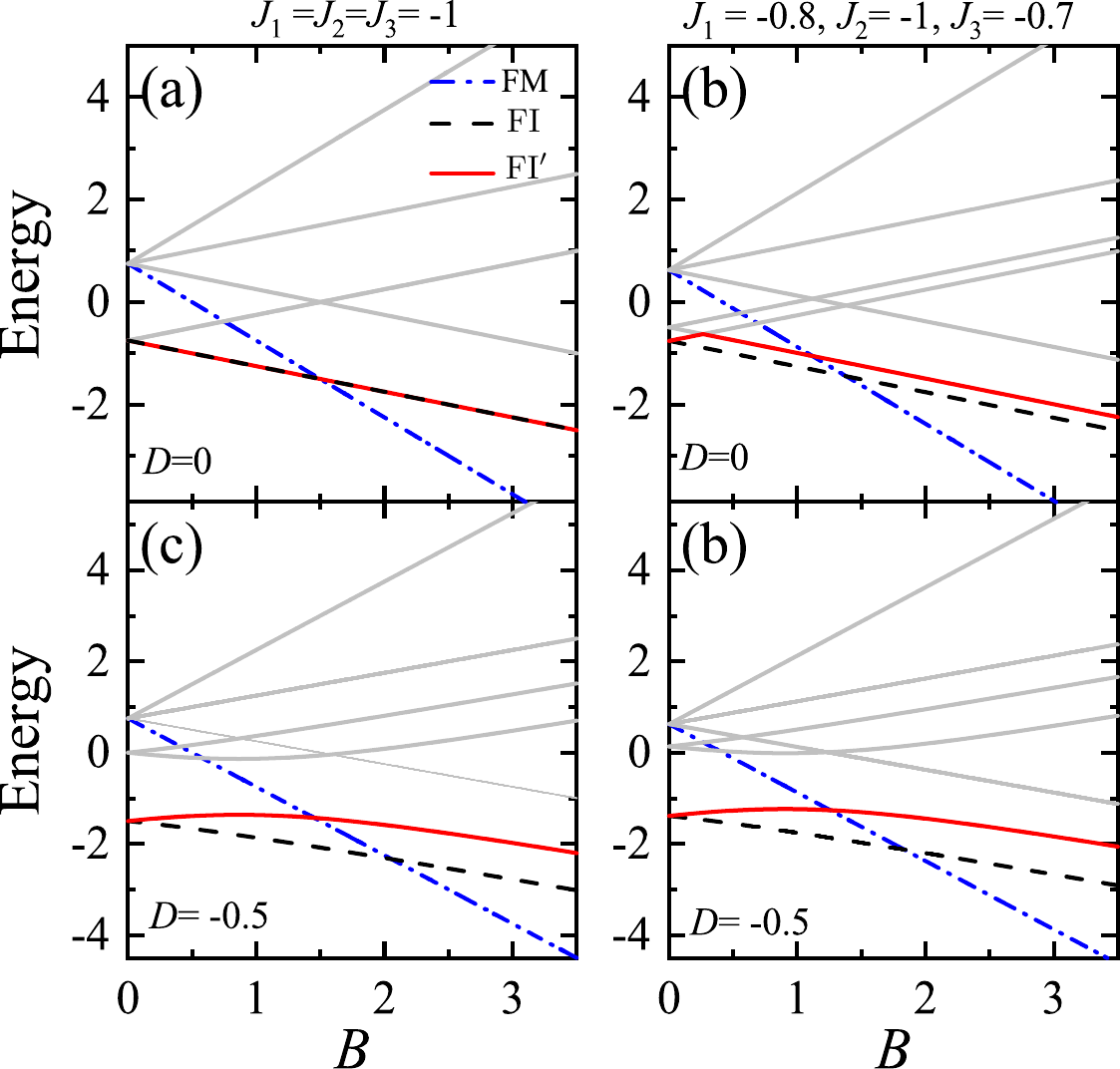} \caption{ (Left column) Energy spectrum as a function of the magnetic field
$B$, for $J_{1}=J_{2}=J_{3}=-1$ and $D=0$. (Right column) Energy
spectrum as a function of $B$ for $J_{1}=-0.8$, $J_{2}=-1$, $J_{3}=-0.7$
and $D=-0.5$. The black, red, and blue curves (as indicated in the
legend) represent the lowest and most relevant energy levels that
determine the ground state, while the gray curves correspond to higher-lying
excitations.}\label{fig:2}
\end{figure}

The ground-state phase diagram provides key insight into the magnetic
behavior of the system. We begin by examining the energy spectrum
shown in Fig.\ref{fig:2}, which displays the energy levels as a function
of the external magnetic field $B$. Two cases are considered: equal
exchange couplings $(J_{1}=J_{2}=J_{3}=J)$ and unequal couplings
$(J_{1}\neq J_{2}\neq J_{3})$, with $D=0$ in the first row and $D=-0.5$
in the second row. The DM interaction $D$ breaks the lattice symmetry
and enhances the magnetic anisotropy of the system. At low temperatures,
the low-lying states dominate the physical behavior; these are indicated
by the black, red, and blue curves in panel (a), while the gray curves
represent higher-energy excitations.

Next, we group the most relevant energy levels according to the total
spin, since this parameter is crucial for determining the magnetic
behavior of the molecule.

There is a unique state in which all spins are aligned with the external
magnetic field, corresponding to a total spin 3/2. This ferromagnetic
(FM) phase, represented by the blue curve in Fig.\ref{fig:2}, is
described by

\begin{align}
|\mathrm{FM}\rangle= & |\uparrow\uparrow\uparrow\rangle,
\end{align}
with ground-state energy

\begin{align}
E_{\mathrm{FM}}= & -\frac{J_{1}}{4}-\frac{J_{2}}{4}-\frac{J_{3}}{4}-\frac{3B}{2},
\end{align}
where the upward arrows denote spins polarized along the $+z$-direction
of the external magnetic field.

Another possible configuration arises when two spins are aligned parallel
and the third is oriented in the opposite direction. Before addressing
the fully general case, let us first examine a particular situation.

The first case corresponds to $J_{1}\ne J_{2}\ne J_{3}$ and $D=0$.
Under these conditions, the system exhibits a ferrimagnetic (FI) phase,
whose ground-state energy is

\begin{align}
E_{\mathrm{FI}}= & \frac{J_{1}}{4}+\frac{J_{2}}{4}+\frac{J_{3}}{4}-\frac{1}{2}\left(B+J_{s}\right),
\end{align}
and the associated ground state is a linear combination of three basis
states,

\begin{alignat*}{1}
|\mathrm{FI}\rangle= & c_{1}|\uparrow\uparrow\downarrow\rangle+c_{2}|\uparrow\downarrow\uparrow\rangle+c_{3}|\downarrow\uparrow\uparrow\rangle,
\end{alignat*}
with coefficients
\begin{alignat*}{1}
c_{1}= & \tfrac{-J_{s}+J_{2}-J_{1}}{\sqrt{\left(J_{s}-J_{2}+J_{1}\right)^{2}+\left(J_{s}-J_{2}+J_{3}\right)^{2}+\left(J_{1}-J_{3}\right)^{2}}},\\
c_{2}= & \tfrac{J_{s}-J_{2}+J_{3}}{\sqrt{\left(J_{s}-J_{2}+J_{1}\right)^{2}+\left(J_{s}-J_{2}+J_{3}\right)^{2}+\left(J_{1}-J_{3}\right)^{2}}},\\
c_{3}= & \tfrac{J_{1}-J_{3}}{\sqrt{\left(J_{s}-J_{2}+J_{1}\right)^{2}+\left(J_{s}-J_{2}+J_{3}\right)^{2}+\left(J_{1}-J_{3}\right)^{2}}}.
\end{alignat*}

The second particular case corresponds to $J_{1}=J_{2}=J_{3}=J$,
with arbitrary $D$ and $B$. In this situation, the energy follows
from \eqref{eq:e6} and reads

\begin{align}
E_{\mathrm{FI}}= & \frac{3J}{4}-\frac{1}{2}K_{+}.\label{eq:EFI1}
\end{align}
The associated ferrimagnetic ground state is
\begin{alignat}{1}
|\mathrm{FI}\rangle= & \tfrac{1}{\sqrt{3(r^{2}_{1}+1)}}\Bigl\{{\rm e}^{\frac{4\pi i}{3}}|\downarrow\downarrow\uparrow\rangle+{\rm e}^{\frac{2\pi i}{3}}|\downarrow\uparrow\downarrow\rangle+|\uparrow\downarrow\downarrow\rangle\nonumber \\
 & r_{1}{\rm e}^{-\frac{3\pi i}{4}}(|\uparrow\uparrow\downarrow\rangle+{\rm e}^{\frac{2\pi i}{3}}|\uparrow\downarrow\uparrow\rangle+{\rm e}^{\frac{\pi i}{3}}|\downarrow\uparrow\uparrow\rangle)\Bigr\},\label{eq:St-FI1}
\end{alignat}
with parameter $r_{1}=\tfrac{\sqrt{3}D+B+K_{+}}{D\sqrt{6}}$. It is
important to note that the eigenstate of the FI phase is independent
of the exchange parameter $J$. The corresponding ferrimagnetic energy
is represented by the black dashed curve in Fig.\ref{fig:2}. Under
the same conditions, the eigenvalue given by Eq.\eqref{eq:e8} (blue
curve in Fig.\ref{fig:2}) has an eigenstate
\begin{alignat}{1}
|\mathrm{FI}'\rangle= & \tfrac{1}{\sqrt{3(r^{2}_{2}+1)}}\Bigl\{{\rm e}^{\frac{2\pi i}{3}}|\downarrow\downarrow\uparrow\rangle+{\rm e}^{\frac{3\pi i}{3}}|\downarrow\uparrow\downarrow\rangle+|\uparrow\downarrow\downarrow\rangle\nonumber \\
 & r_{2}{\rm e}^{\frac{\pi i}{4}}(|\uparrow\uparrow\downarrow\rangle+{\rm e}^{\frac{\pi i}{3}}|\uparrow\downarrow\uparrow\rangle+{\rm e}^{\frac{2\pi i}{3}}|\downarrow\uparrow\uparrow\rangle)\Bigr\},\label{eq:St-FI2}
\end{alignat}
with $r_{2}=\tfrac{-\sqrt{3}D+B+K_{-}}{D\sqrt{6}}$. The state $|\mathrm{FI}'\rangle$
does not constitute an independent phase, except for the case of negative
magnetic field $B$.

Particular attention must be paid when the exchange couplings satisfy
$J_{1}=J_{2}=J_{3}$. In this case, the molecule exhibits geometric
frustration due to the triangular arrangement of antiferromagnetically
coupled spins, which prevents all pairwise interactions from being
simultaneously minimized.

A particularly relevant situation occurs when $J_{1}=J_{2}=J_{3}$
and $D\to0$. In this limit, the two lowest-energy states become degenerate,
forming a doublet. In the following, we refer to this regime as FR,
emphasizing its connection with the frustrated triangular geometry.
For the eigenstates of Eqs. \eqref{eq:St-FI1} and\eqref{eq:St-FI2},
the limit $D\rightarrow0$leads to $r_{1}\rightarrow\infty$ and $r_{2}\rightarrow\infty$.
In this limit, the first three terms in both eigenstates vanish, and
the normalized eigenstates of this degenerate doublet reduce to 
\begin{equation}
|\mathrm{FR}\rangle=\begin{cases}
\tfrac{1}{\sqrt{3}}\left(|\uparrow\uparrow\downarrow\rangle+{\rm e}^{\frac{2\pi i}{3}}|\uparrow\downarrow\uparrow\rangle+{\rm e}^{\frac{\pi i}{3}}|\downarrow\uparrow\uparrow\rangle\right),\\
{\rm or}\\
\tfrac{1}{\sqrt{3}}\left(|\uparrow\uparrow\downarrow\rangle+{\rm e}^{\frac{\pi i}{3}}|\uparrow\downarrow\uparrow\rangle+{\rm e}^{\frac{2\pi i}{3}}|\downarrow\uparrow\uparrow\rangle\right).
\end{cases}
\end{equation}
The frustrated (FR) regime discussed here is therefore directly connected
to the trigonal symmetry of the equilateral triangle and to the associated
doubly degenerate E representation. The residual entropy $\mathcal{S}=\ln2$
originates from the degeneracy of the two chiral components. Both
the DM interaction $\vec{D}$ and asymmetric exchange couplings $J_{1}\neq J_{2}\neq J_{3}$
lower the trigonal symmetry, lift this degeneracy, and suppress the
FR regime, as illustrated below in the phase diagrams and entropy
curves.

For arbitrary Hamiltonian parameters {[}\eqref{ham}{]}, the eigenstate
corresponding to the ferrimagnetic phase can in general be expressed
as
\begin{align}
|\mathrm{FI}\rangle= & c_{1}|\uparrow\uparrow\downarrow\rangle+c_{2}|\uparrow\downarrow\uparrow\rangle+c_{3}|\downarrow\uparrow\uparrow\rangle\nonumber \\
 & +c_{4}|\downarrow\downarrow\uparrow\rangle+c_{5}|\downarrow\uparrow\downarrow\rangle+c_{6}|\uparrow\downarrow\downarrow\rangle,
\end{align}
where the coefficients $c_{i}$ are, in general, complex amplitudes
satisfying the normalization condition $\sum^{6}_{j=1}|c_{j}|^{2}=1$.

The general ferrimagnetic state has energy
\begin{align}
E_{\mathrm{FI}}= & \frac{J_{1}}{4}+\frac{J_{2}}{4}+\frac{J_{3}}{4}-\frac{1}{2}K_{+}.
\end{align}

From Fig.\ref{fig:2}, it is evident that the ferromagnetic (FM) phase
becomes stable under strong external magnetic fields, independent
of the exchange couplings. The system undergoes a first-order phase
transition from the FI to the FM state as the magnetic field increases.
The critical field for this transition is
\begin{equation}
B_{c}=\frac{d-3J_{s}}{8}+\frac{1}{8}\sqrt{\left(5d-J_{t}\right)^{2}+4J_{t}d-16J^{2}_{s}},
\end{equation}
with
\begin{align*}
d= & \sqrt{3D^{2}+J^{2}_{s}},\\
J_{t}= & J_{1}+J_{2}+J_{3},
\end{align*}

At this transition, the total spin of the molecule increases from
$S=1/2$ (FI phase) to $S=3/2$ (FM phase), in agreement with theoretical
expectations.

This behavior is illustrated in Fig.\ref{fig:2}(a), where the double
degeneracy in energy persists up to the critical magnetic field $B_{c}$.
A degenerate regime emerges when the two lowest-energy states become
equivalent. At the critical field $B_{c}=1.5$, a phase transition
takes place from the FR phase to the FM phase. In Fig.\ref{fig:2}(b),
the energy levels are shown for unequal exchange parameters; in this
case, the FR degeneracy is lifted, and only the FI phase remains.
A similar situation is observed in panel (c), where frustration is
removed by the DM interaction. Finally, panel (d) displays energy
levels very similar to those in panel (c).

It is instructive to relate the present results to available experimental
data on triangular $\mathrm{Cu}_{3}$ clusters. Inelastic neutron
scattering measurements report low-energy excitations of the order
of 0.5-0.6 meV, together with an additional splitting around 0.1 meV
attributed to DM interactions \citep{lida}. In our model, the overall
energy scale is set by the exchange parameters $J_{i}$, while the
DM interaction controls the splitting of low-energy levels. By choosing
parameters in the same energy range, the model reproduces qualitatively
the characteristic hierarchy of energy scales observed experimentally,
namely a dominant exchange energy and a smaller DM-induced splitting.
This supports the interpretation that the main features of the spectrum
and the associated thermodynamic behavior are governed by the interplay
between exchange interactions and antisymmetric exchange terms.

\begin{figure}[h]
\includegraphics[scale=0.45]{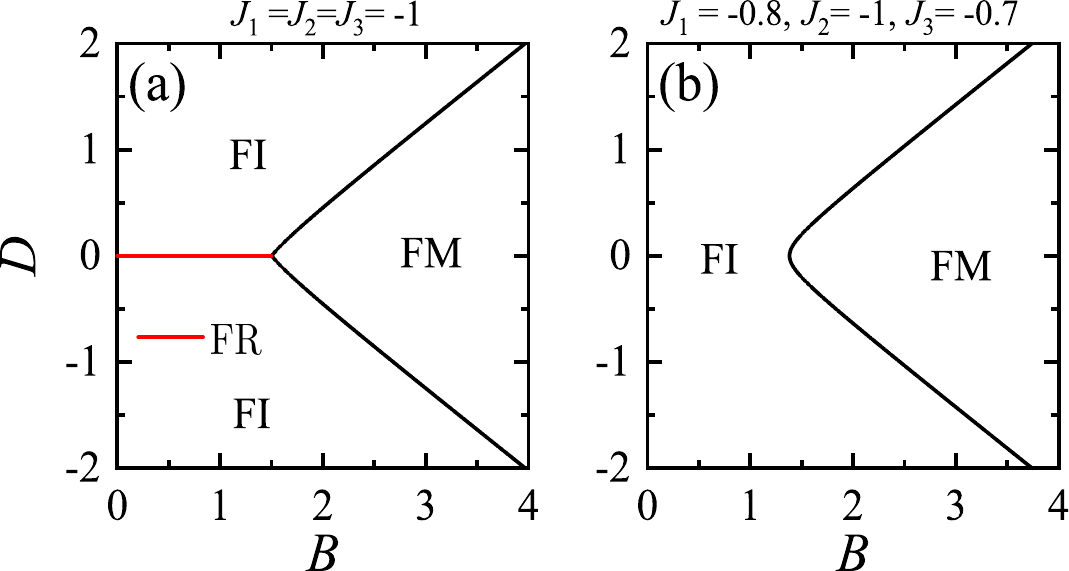}\caption{(a) Zero-temperature phase diagram in the $B-D$ plane for $J_{1}=J_{2}=J_{3}=-1$.
(b) Phase diagram for $J_{1}=-0.8$, $J_{2}=-1$, and $J_{3}=-0.7$.
The black solid line marks the transition between the FI and FM phases,
while the red line indicates the region with a twofold degenerate
ground state (denoted FR).}\label{fig:3}
\end{figure}

To gain deeper insight into the quantum phase transitions, we now
analyze the ground-state phase diagram. Fig.\ref{fig:3}(a) shows
the phase diagram in the $B\lyxmathsym{\textendash}D$ plane for the
symmetric case $J_{1}=J_{2}=J_{3}=-1$. In this situation, the system
exhibits two stable ground states: a ferrimagnetic (FI) phase and
a ferromagnetic (FM) phase. The red line corresponds to a frustrated
(FR) phase, characterized by a residual entropy $\mathcal{S}=\ln(2)$.
For a symmetric triangle with vanishing DM interaction ($D=0$), the
FI and $\mathrm{FI}'$ states are degenerate for $B<B_{c}$, as seen
in Fig.\ref{fig:2}(a). This degeneracy leads to a twofold ground
state, resulting in a residual entropy$\mathcal{S}=\ln2$. At the
critical magnetic field $B_{c}=1.5$, at the transition field, the
FI, $\mathrm{FI}'$, and FM states become degenerate, leading to a
residual entropy of $\mathcal{S}=\ln(3)$.

Fig.\ref{fig:3}(b) presents the phase diagram for asymmetric couplings,
with $J_{1}=-0.8$, $J_{2}=-1$, and $J_{3}=-0.7$. In this case,
the frustration disappears due to the lack of symmetry in the exchange
interactions, and the residual entropy at the phase boundary is reduced
to $\mathcal{S}=\ln(2)$.

\subsection{Magnetization and Magnetic Susceptibility}

The magnetization of the Heisenberg model on a triangular lattice
offers direct insight into the underlying exchange interactions and
the resulting magnetic behavior. It constitutes a key quantity for
characterizing the magnetic properties and associated phenomena in
triangular molecular systems.

\begin{figure}[h]
\includegraphics[scale=0.6]{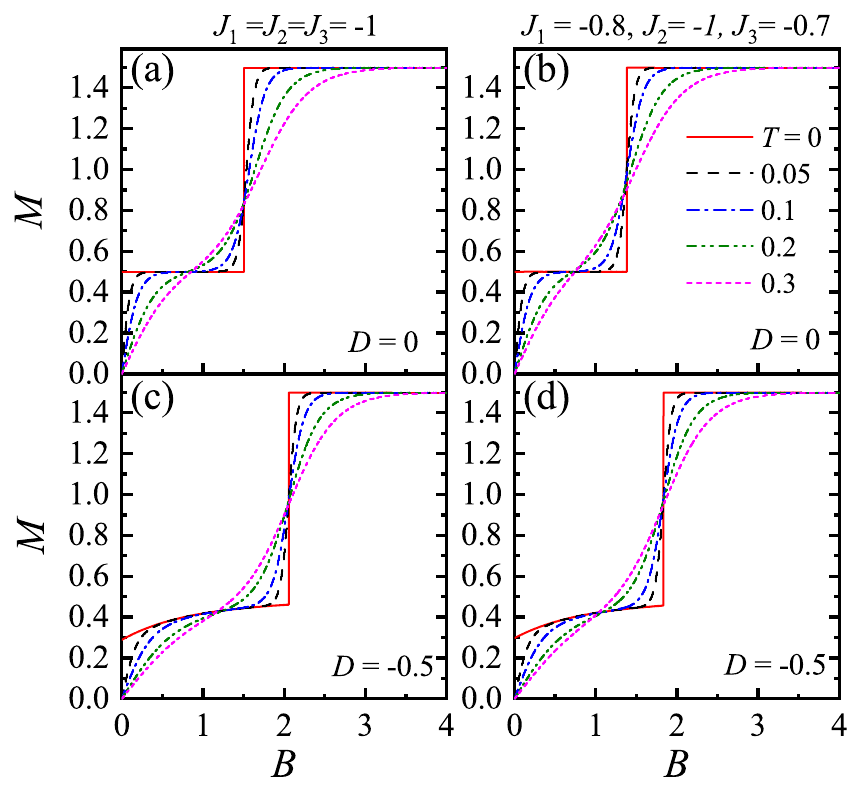} \caption{ Total magnetization $M$ as a function of the magnetic field $B$
for different temperatures and values of $D$. (Left column) Results
for $J_{1}=J_{2}=J_{3}=-1$. (Right column) Results for $J_{1}=-0.8$,
$J_{2}=-1$, and $J_{3}=-0.7$.}\label{fig:4}
\end{figure}

Figure \ref{fig:4} shows the total magnetization $M$ as a function
of the external magnetic field $B$ for different temperatures $T$,
with the values of $T$ indicated in panel (b). At zero temperature,
the system exhibits a finite magnetization, which gradually decreases
as $T$ increases due to the thermal destruction of spin order. This
behavior reflects the fact that, although the system is a finite quantum
cluster (zero-dimensional), its thermal behavior resembles that of
a finite classical spin system without finite-temperature phase transitions.
At finite temperatures, the ground-state 1/3-magnetization plateau
becomes progressively smoother and eventually disappears at high fields,
as illustrated in panels (a) and (b). In contrast, for $T=0$, panels
(c) and (d) do not display the 1/3 plateau because of the presence
of the DM interaction, which induces a nontrivial field dependence
of the magnetization. For finite temperatures, however, the magnetization
curves become smoother, resembling the behavior observed for $D=0$.

\begin{figure}[h]
\includegraphics[scale=0.6]{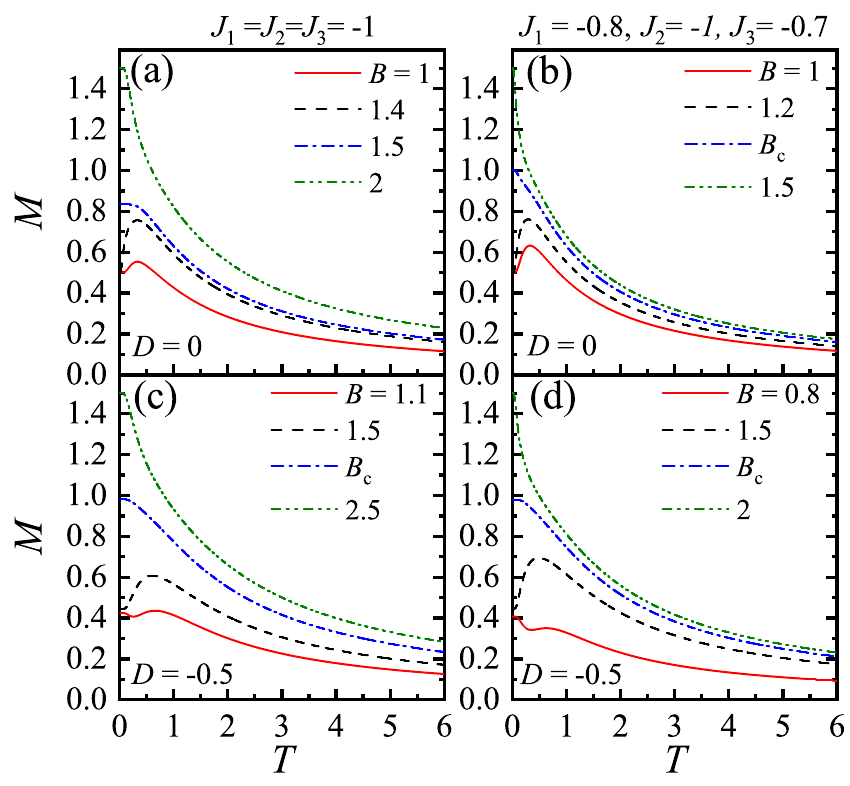} \caption{Total magnetization $M$ as a function of temperature $T$ for different
values of magnetic field $B$ and $D$, with $J_{1}=J_{2}=J_{3}=-1$.
(Right column) Corresponding results for $J_{1}=-0.8$, $J_{2}=-1$,
and $J_{3}=-0.7$.}\label{fig:5}
\end{figure}

The temperature dependence of the total magnetization is shown in
Fig.\ref{fig:5}. In the left column, $M(T)$ is plotted for four
different values of the external magnetic field $B$. For fields below
the critical value $B_{c}$, the magnetization displays an anomalous
maximum: it increases at low temperatures, reaches a peak, and then
decreases monotonically as thermal fluctuations become stronger, progressively
misaligning the spins and driving the magnetization to zero at sufficiently
high temperatures. This nonmonotonic behavior reflects the competition
between spin alignment induced by the field and disorder induced by
thermal agitation. At the critical field, the magnetization remains
nearly constant at low temperatures, but decreases monotonically once
thermal effects dominate. For strong magnetic fields, by contrast,
the magnetization is a strictly decreasing function of temperature.

\begin{figure}[h]
\includegraphics[scale=0.6]{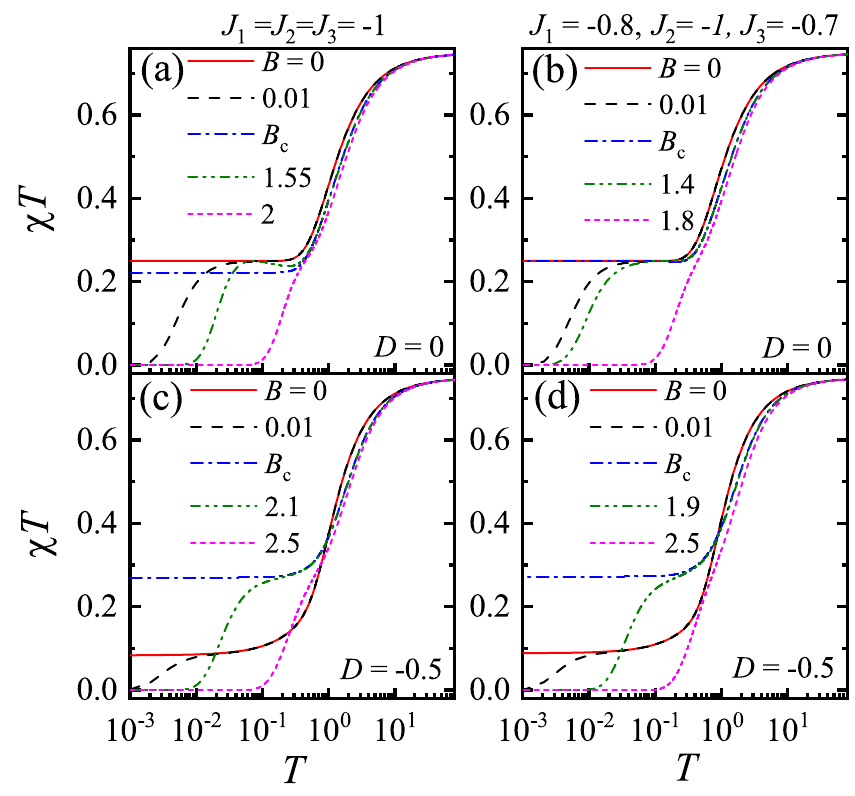} \caption{Magnetic susceptibility times temperature, $\chi T$, as a function
of temperature $T$ (logarithmic scale). (Left column) Results for
$J_{1}=J_{2}=J_{3}=-1$, with different values of $B$ and $D$. (Right
column) Results for $J_{1}=-0.8$, $J_{2}=-1$, and $J_{3}=-0.7$,
for several $B$ and $D$.}\label{fig:6}
\end{figure}

Figure \ref{fig:6} shows the magnetic susceptibility times temperature,
$\chi T$, as a function of temperature $T$ on a logarithmic scale.
We plot $\chi T$ rather than $\chi$ itself in order to emphasize
the low-temperature behavior, where $\chi$ diverges as $1/T$. In
panel (a), for $B=0$ we find $\chi T=0.25$ at low temperatures,
while at the critical field $\chi T\approx0.222$, corresponding to
$\chi\to\infty$. For all other field values, $\chi T\to0$ as $T\to0$,
indicating that $\chi$ remains finite. Panel (b) displays a qualitatively
similar behavior: for both $B=0$ and the critical field, $\chi T=0.25$.
In panel (c), the low-temperature values are $\chi T\approx0.083$
at $B=0$ and $\chi T\approx0.269$ at the critical field, while in
panel (d) they are $\chi T\approx0.0883$ and $\chi T\approx0.272$,
respectively. At high temperatures, and for all values of the magnetic
field, $\chi T$ approaches the Curie-law behavior characteristic
of a paramagnetic material, as expected.

\subsection{Entropy and specific heat}

Continuing with our discussion, we will now explore entropy, which
plays an important role in studying the magnetic behavior of Heisenberg
models on triangular structures, particularly in relation to ground-state
phase transitions. At low temperatures, the entropy is governed by
the degeneracy of the lowest-energy states, so that only a subset
of the total Hilbert space may contribute.

\begin{figure}[h]
\includegraphics[scale=0.6]{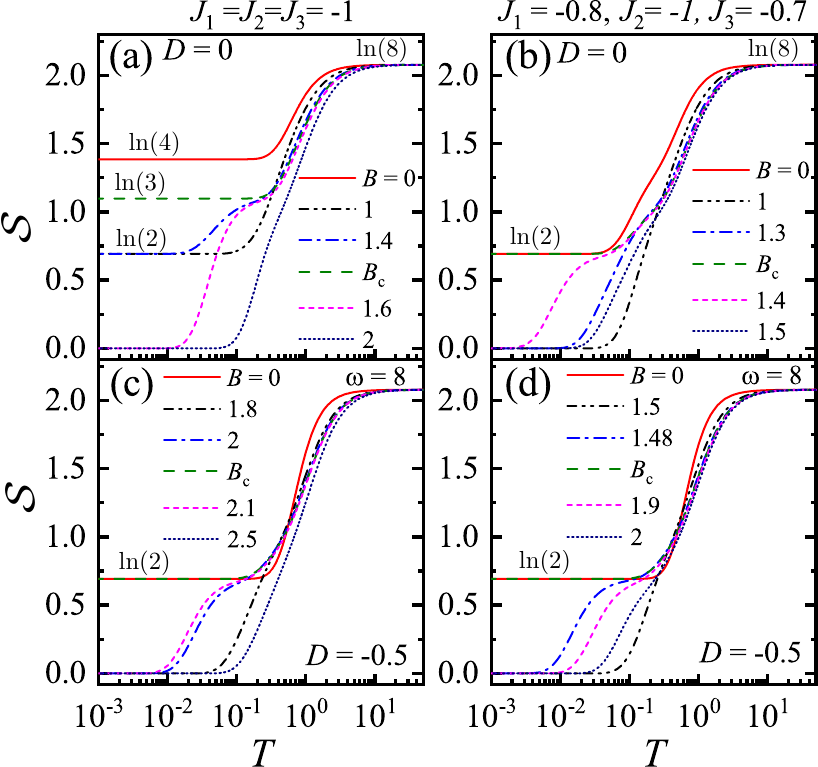} \caption{ Entropy $\mathcal{S}$ as a function of temperature $T$ (logarithmic
scale). (Left column) Results for $J_{1}=J_{2}=J_{3}=-1$, with different
values of $B$ and $D$. (Right column) Results for $J_{1}=-0.8$,
$J_{2}=-1$, and $J_{3}=-0.7$, for various values of $B$ and $D$.}\label{fig:7}
\end{figure}

Figure \ref{fig:7} shows the entropy $\mathcal{S}$ as a function
of temperature for a range of values of the magnetic field $B$. At
low temperatures, several residual entropy plateaus are observed,
reflecting degeneracies of the low-energy spectrum. In panel (a),
for $B=0$, four states are degenerate, yielding a residual entropy
$\mathcal{S}=\ln(4)$, consistent with Fig.\ref{fig:2}. At the critical
field, $\mathcal{S}=\ln(3)$, reflecting the coexistence of the FM,
FI, and $\mathrm{FI}'$ states. For $0<B<B_{c}$, a twofold degenerate
ground state appears, giving rise to a residual entropy $\mathcal{S}=\ln(2)$.
For $B>B_{c}$, no residual entropy remains (see Fig. \ref{fig:3}).
Panel (b) exhibits residual entropy $\mathcal{S}=\ln(2)$ at both
$B=0$ and $B=B_{c}$; for all other field values, no residual entropy
is present due to the asymmetric exchange couplings. Panel (c) shows
a similar trend: although the exchange interactions are symmetric,
the presence of $D=-0.5$ breaks the symmetry and eliminates the $\mathcal{S}=\ln(2)$
plateau seen in panel (a) for $0<B<B_{c}$. Panel (d) displays behavior
analogous to panels (b) and (c). At high temperatures, the entropy
approaches the expected value $\mathcal{S}=\ln(8)$, since in this
limit all eight states are equally probable, yielding the maximum
entropy.

\begin{figure}[h]
\includegraphics[scale=0.41]{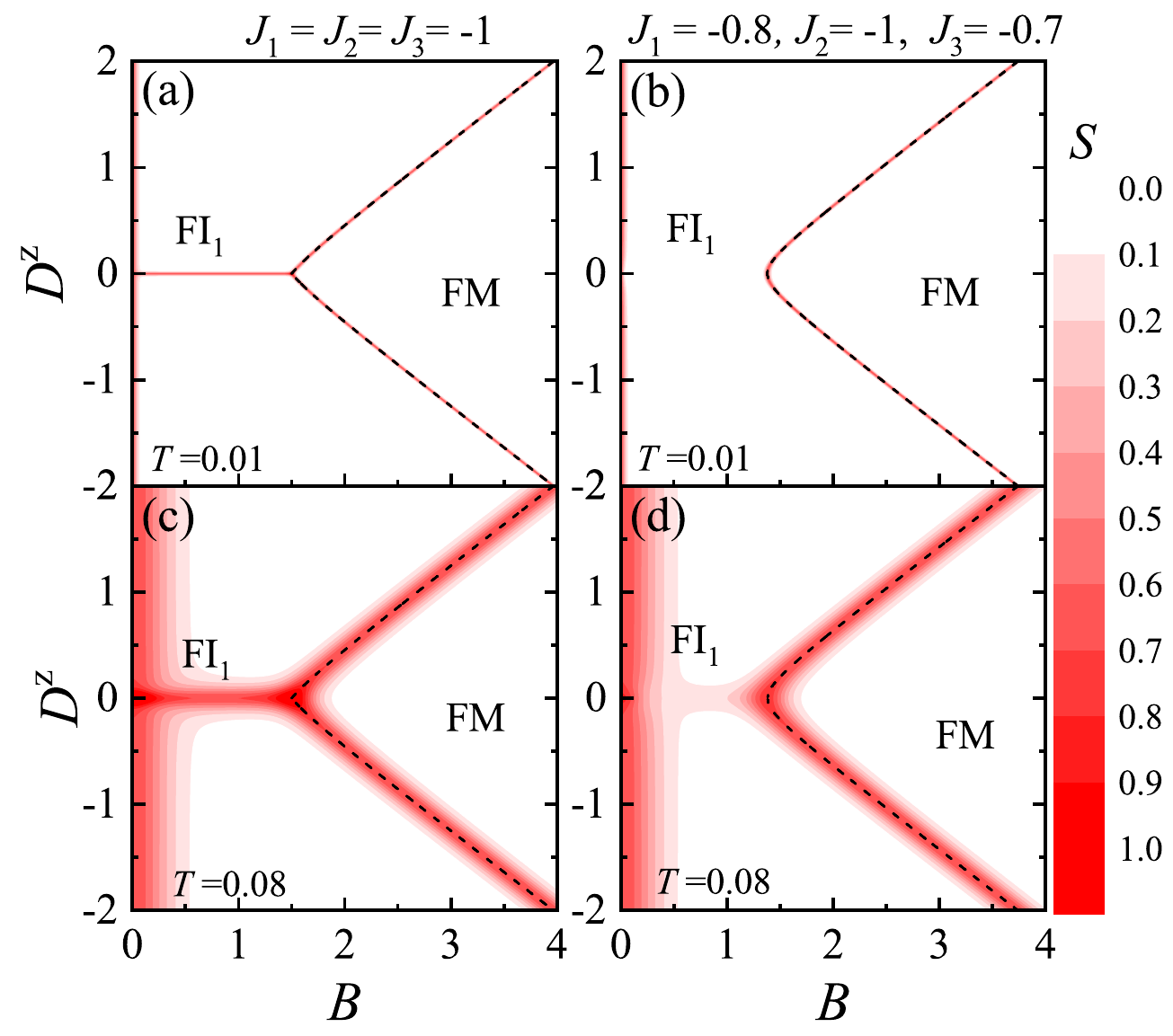} \caption{(Left column) Density plots of the entropy $\mathcal{S}$ as a function
of $D$ and $B$ for $J_{1}=J_{2}=J_{3}=-1$. Results are shown for
two temperatures, $T=0.01$ and $T=0.08$. The color scale on the
right indicates the entropy values. (Right column) Corresponding plots
for $J_{1}=-0.8$, $J_{2}=-1$, and $J_{3}=-0.7$, also at $T=0.01$
and $T=0.08$.}\label{fig:8}
\end{figure}

Figure \ref{fig:8} shows density plots of the entropy in the $B\lyxmathsym{\textendash}D$
plane, with the red color scale indicating the entropy magnitude.
Panel (a) presents the case $T=0.01$, superimposed on the zero-temperature
phase diagram. The nearly red line coincides with the phase boundaries,
demonstrating how finite-temperature entropy reflects the underlying
ground-state transitions. Panel (c) shows the corresponding plot at
a higher temperature, $T=0.08$. Here, a highly degenerate region
appears due to the coexistence of several low-energy states, yielding
$\mathcal{S}=\ln(3)$ is accompanied by a pronounced red line, highlighting
the role of degeneracy. Enhanced entropy also appears at $B=0$ and
along the FI-FM phase boundary. Panel (b) displays the entropy at
$T=0.01$ for the asymmetric case, $J_{1}=-0.8$, $J_{2}=-1$, and
$J_{3}=-0.7$. The entropy still follows the zero-temperature phase
structure. Panel (d) shows the results for $T=0.08$, qualitatively
similar to panel (c) but without the frustrated region for $0<B<B_{c}$,
as expected from the broken symmetry of the exchange couplings.

\begin{figure}[h]
\includegraphics[scale=0.58]{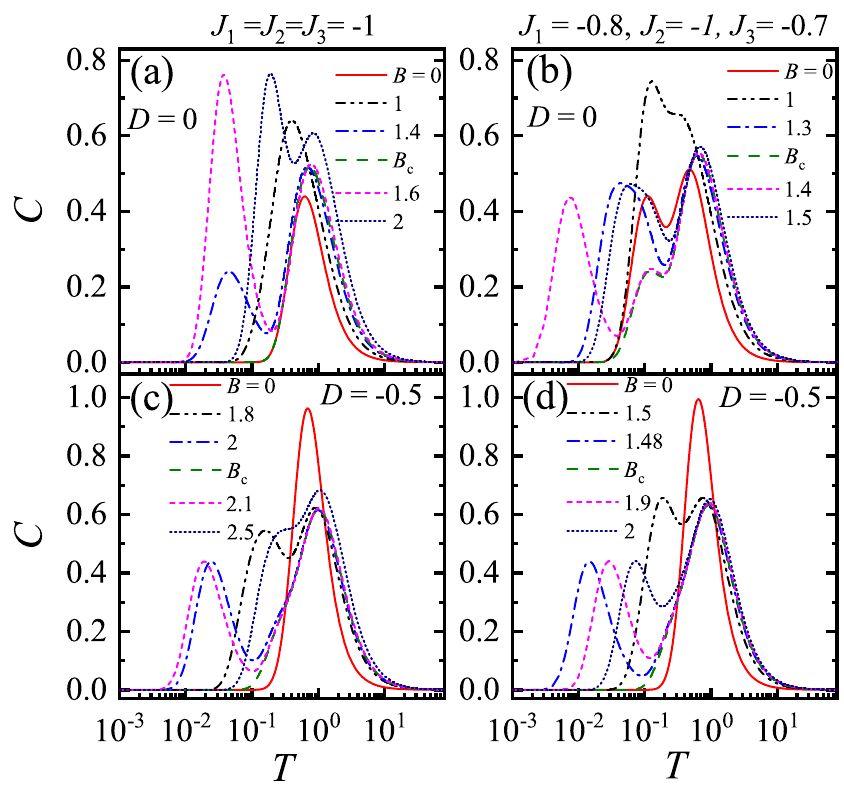} \caption{Specific heat $C$ as a function of temperature $T$ (logarithmic
scale). (Left column) Results for $J_{1}=J_{2}=J_{3}=-1$, with two
values of $D$ and several values of $B$. (Right column) Results
for $J_{1}=-0.8$, $J_{2}=-1$, and $J_{3}=-0.7$, also for two values
of $D$ and several values of $B$.}\label{fig:9}
\end{figure}

The specific heat of the Heisenberg model on a triangular structure
is an important quantity, as it provides information about thermal
behavior, phase transitions, and magnetic excitations. Because the
energy spectrum of the system is finite, the specific heat is expected
to display Schottky-type behavior, vanishing as $T\to0$ in accordance
with thermodynamic principles. This behavior is confirmed in Fig.\ref{fig:9}(a),
which shows $C(T)$ for different values of the magnetic field. A
typical Schottky peak appears, associated with the change in entropy
curvature, since entropy increases monotonically with temperature.
The Schottky anomaly persists even when the parameters are tuned to
the zero-temperature phase transition. In the low-temperature region,
the curves for $B=0$ and $B=B_{c}$ coincide, but they separate at
higher temperatures. For $B\lesssim B_{c}$, the anomalous peak is
suppressed relative to the Schottky peak, whereas for $B\gtrsim B_{c}$,
the anomalous peak becomes more pronounced. Panel (b) shows the coexistence
of the anomalous peak with the standard Schottky peak; in this case,
the anomalous feature is present even for $B=0$ and $B=B_{c}$, and
the overall specific heat differs significantly from panel (a). Panel
(c) exhibits qualitatively similar behavior to panel (a), although
here the specific heat curves for $B=0$ and $B=B_{c}$ deviate noticeably.
Finally, panel (d) displays results similar to those of panel (c).

\subsection{Magnetocaloric effect}

We now turn to the MCE in the Heisenberg model on a triangular structure.
Studying the MCE is crucial for understanding the coupled magnetic
and thermodynamic behavior of triangular molecular systems containing
$\mathrm{Cu}^{2+}$ ions. In particular, the MCE reveals how magnetic
interactions and temperature variations interplay, offering valuable
insight into the physics of frustrated systems and pointing toward
potential applications in magnetic refrigeration and energy-efficient
cooling technologies.

\begin{figure}[h]
\includegraphics[scale=0.5]{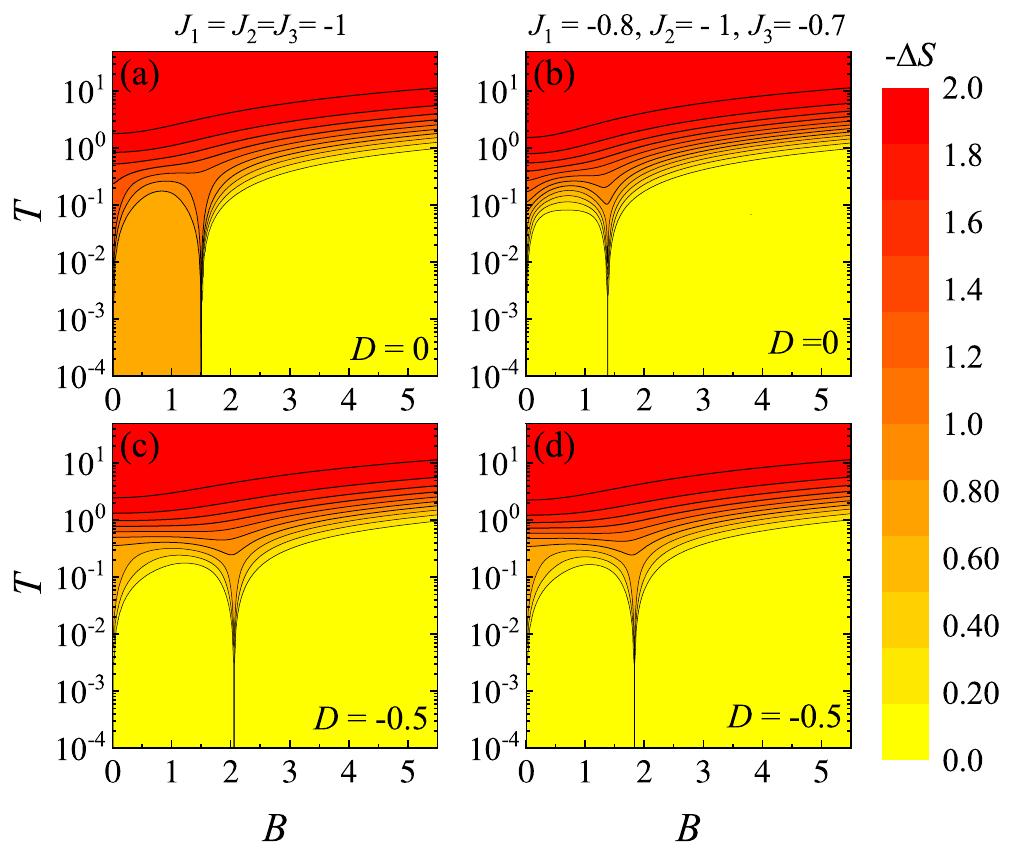} \caption{Density plots of the entropy $\mathcal{S}$ in the $B-T$ plane (logarithmic
temperature scale). The color scale on the right indicates the entropy
magnitude. (Left column) Results for $J_{1}=J_{2}=J_{3}=-1$, with
two values of $D$. (Right column) Results for $J_{1}=-0.8$, $J_{2}=-1$,
and $J_{3}=-0.7$, also with two values of $D$.}\label{fig:10}
\end{figure}

The adiabatic behavior of the triangular-shaped molecule can be analyzed
from Fig.\ref{fig:10}(a). This panel shows the entropy density, represented
by the color gradient scale, as a function of magnetic field $B$
and temperature $T$ (logarithmic scale) for different values of $D$.
The contour lines correspond to adiabatic variations of entropy with
temperature and field. A pronounced depression of $T$ as a function
of $B$ appears at the critical field $B_{c}=1.5$. For $0<B<B_{c}$,
an orange region marks the frustrated FR phase with residual entropy
$\mathcal{S}=\ln(2)$, surrounded by a more intense region where $\mathcal{S}\approx\ln(3)$.
This behavior is consistent with Fig.\ref{fig:2}(a) and Fig.\ref{fig:7}.
In contrast, panels (b)-(d) show qualitatively similar adiabatic curves,
again manifesting as depressions in the logarithmic scale, which coincide
with the zero-temperature phase transitions. In these cases, the yellow
regions are bordered by orange regions, primarily reflecting residual
entropy $\mathcal{S}=\ln(2)$. Another important quantity is the magnetic
entropy variation, $-\Delta\mathcal{S}$, as defined in Eq.\eqref{mce}.
This measure is central to the analysis of the MCE in the Heisenberg
triangular model, as it directly reflects field-induced changes in
magnetic order and disorder. Understanding and controlling this entropy
variation may be crucial for the design of efficient magnetic refrigeration
systems and for advancing our knowledge of magnetism in frustrated
triangular arrangements.

\begin{figure*}
\includegraphics{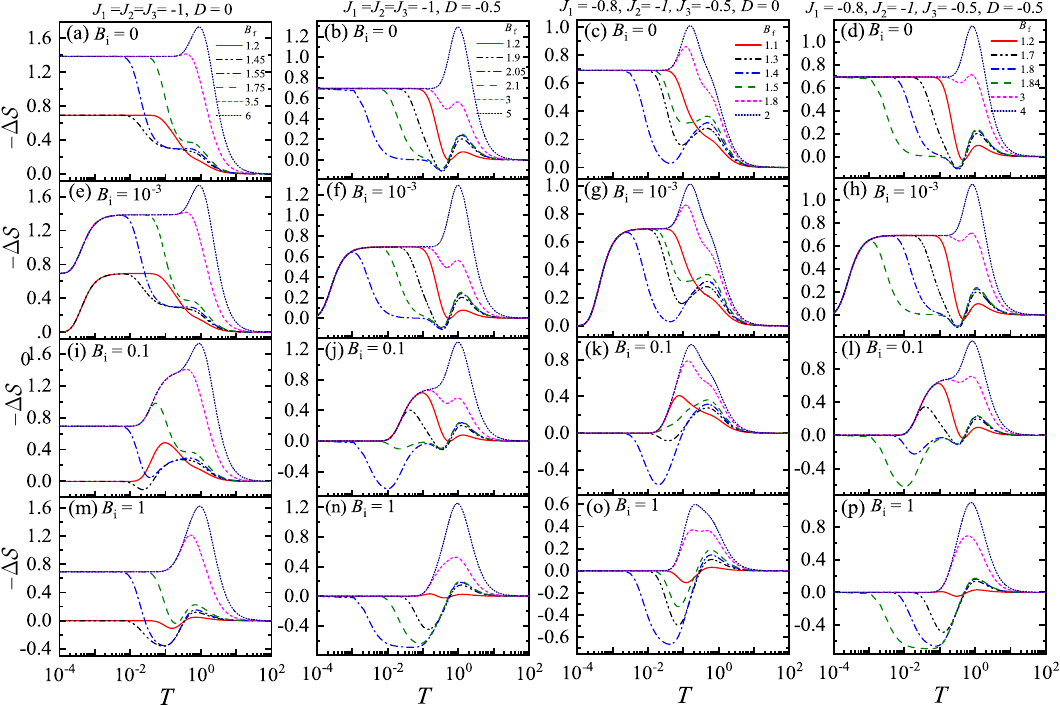} \caption{ Isothermal entropy variation, $-\Delta\mathcal{S}$, as a function
of temperature $T$ (logarithmic scale) for several final magnetic
fields $B_{f}$, starting from different initial fields. The first
row corresponds to $B_{i}=0$, the second to $B_{i}=10^{-3}$, the
third to $B_{i}=0.1$, and the fourth to $B_{i}=1$. The first and
third columns show results for $D=0$, while the second and fourth
columns correspond to $D=-0.5$. The values of the exchange couplings
are indicated at the top of each column.}\label{fig:11}
\end{figure*}

The isothermal entropy change, $-\Delta\mathcal{S}$, as a function
of temperature $T$, is shown in Fig. \ref{fig:11} for several final
magnetic fields $B_{f}$, as indicated in panels (a-b). The first
and third columns correspond to $D=0$, while the second and fourth
columns correspond to $D=-0.5$. At high temperatures, $-\Delta\mathcal{S}$
vanishes for all values of $B_{i}$ and $B_{f}$, since in this limit
the entropy is the same for all fields (see Fig.\ref{fig:7}). In
the first row, we observe that for large $B_{f}$, small plateaus
in $-\Delta\mathcal{S}$ develop into peaks, which become broader
and more pronounced as the field increases.

All curves in panels (a-d) show a finite $-\Delta\mathcal{S}$ at
$T=0$, a direct consequence of the ground-state degeneracy at $B_{i}=0$.
When $B_{i}\neq0$, the degeneracy is lifted, and $-\Delta\mathcal{S}$
vanishes at low temperatures, except in the symmetric case where FI
and $\mathrm{FI}'$ remain degenerate {[}see Fig.\ref{fig:2}(a){]}.
In panels (a), (c), (e), and (g) ($D=0$), a positive entropy change
is observed, indicating that the system exhibits only the direct magnetocaloric
effect, as $\mathcal{S}(B_{i}=0)>\mathcal{S}$ ($B_{f}\neq0$). However,
for $B_{i}\geq0.1$ {[}see panels (i), (k), (m), and (o){]}, the system
begins to display the inverse MCE. For $D=-0.5$ (second and fourth
columns), both direct and inverse effects appear, with the inverse
MCE becoming more pronounced when $B_{i}\geq0.1$.

A characteristic feature of the entropy variation is that $-\Delta\mathcal{S}$
often develops a minimum followed by a peak at intermediate fields,
typically near the phase transition (except in panel a). The minimum
originates from the fact that, for finite fields, the entropy rises
earlier than for $B=0$; in some cases, the corresponding curves even
cross, as illustrated in Fig. \ref{fig:7}. For $D\neq0$, this effect
becomes particularly pronounced: panels (c-f) reveal negative values
of the entropy variation, signaling the emergence of the inverse MCE.

\begin{figure}[h]
\includegraphics[scale=0.5]{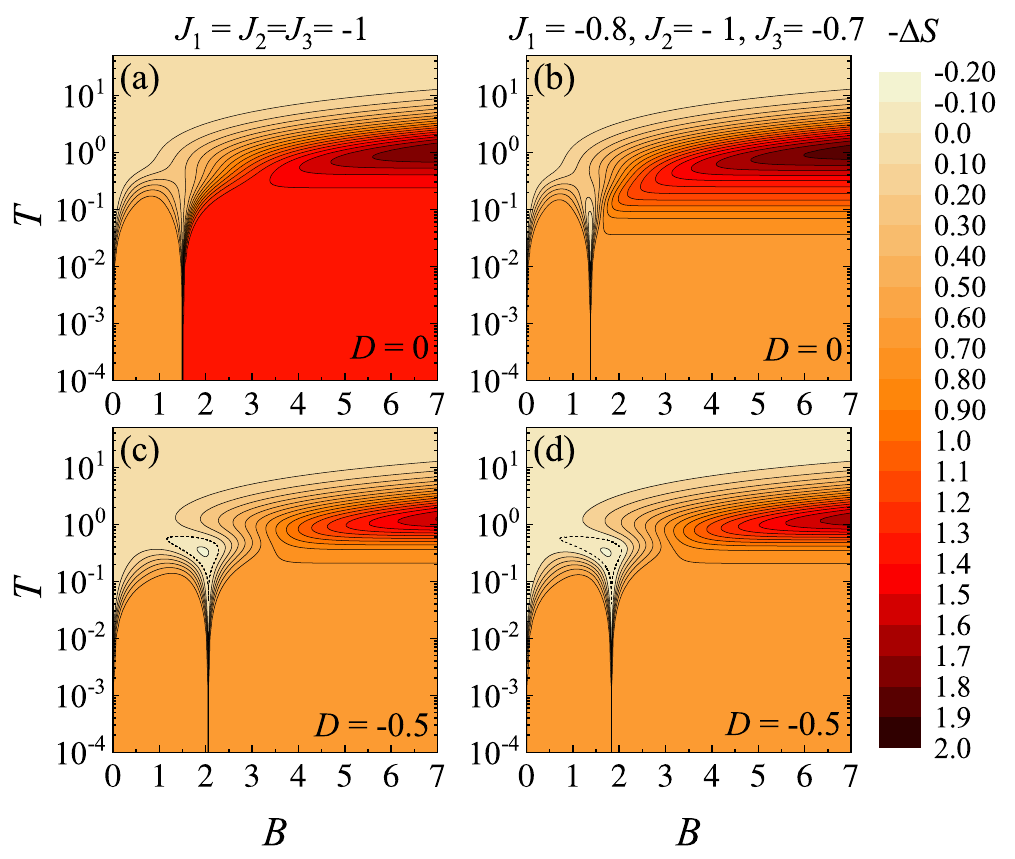} \caption{ Density plots of the magnetic entropy variation, $-\Delta\mathcal{S}$,
in the $B-T$ plane (logarithmic temperature scale) for two values
of $D$. The color scale on the right indicates the magnitude of $-\Delta\mathcal{S}$.
The dashed line marks the locus of zero entropy variation. The exchange
coupling values are specified at the top of the figure.}\label{fig:15}
\end{figure}

Figure \ref{fig:15} shows the entropy variation in the $B-T$ plane
(logarithmic scale), with several contours of constant magnetic entropy
change also plotted. In panel (a), an orange region appears for $0<B<B_{c}$
up to $T\approx0.1$, corresponding to a nearly constant entropy variation
of $-\Delta\mathcal{S}\approx\ln(2)$. For higher fields ($B>B_{c}$),
a red region emerges with $-\Delta\mathcal{S}\approx\ln(4)$. This
behavior arises because the initial entropy is $\mathcal{S}_{i}\approx\ln(4)$,
while the final entropy approaches $\mathcal{S}_{f}\to0$, as the
system becomes ferromagnetically ordered. In panel (b), we again observe
$-\Delta\mathcal{S}\approx\ln(2)$ at low temperatures. At $B_{c}$
and $T\approx0.1$, only a small entropy variation occurs, with $-\Delta\mathcal{S}>0$,
and no inverse MCE is present. In contrast, panels (c) and (d) display
the emergence of the inverse MCE, highlighted by the dashed contour
line.

\section{Concluding remarks}\label{sec:Clsn}

In this work, we have investigated the magnetic and thermodynamic
properties of a triangular spin-1/2 cluster composed of $\mathrm{Cu}^{2+}$-like
magnetic ions. Our analysis encompassed the exact energy spectrum,
ground-state phase diagram, and thermodynamic properties of the Heisenberg
triangle with DM interactions. We have clarified how exchange anisotropy
and DM terms tune the interplay between geometric frustration, degeneracy
of low-energy states, and the MCE, with both direct and inverse regimes
emerging. These findings highlight the potential of triangular single-molecule
magnets for nanoscale refrigeration.

The energy spectrum was studied as a function of the magnetic field
for different exchange interaction values. We found that the low-energy
states dominate the low-temperature regime, thereby determining the
ground-state phase. The phase diagram revealed two stable phases:
a ferrimagnetic (FI) phase and a ferromagnetic (FM) phase. In addition,
a frustrated (FR) phase was identified as a special case when $J_{1}=J_{2}=J_{3}$
and $D=0$, persisting up to a critical magnetic field. A first-order
phase transition from FI to FM occurs upon increasing the magnetic
field.

The magnetization as a function of field and temperature exhibited
anomalous low-temperature peaks, while at high temperatures it displayed
typical paramagnetic behavior. The magnetic susceptibility showed
similar features, with peaks and plateaus at low temperatures for
$D=0$. In contrast, for $D=-0.5$, the $1/3$-plateau vanished due
to the DM interaction. The entropy and specific heat revealed residual
entropy plateaus at low temperatures, reflecting ground-state degeneracies,
while at high temperatures the entropy approached the expected value,
consistent with maximum disorder. The specific heat exhibited Schottky-type
peaks and additional anomalous features near critical fields.

The magnetocaloric effect was also analyzed. Both direct and inverse
MCE were observed depending on the choice of initial and final magnetic
fields. The entropy variation displayed characteristic minima followed
by peaks for intermediate fields, with the inverse MCE becoming more
pronounced at larger initial fields.

In summary, our results provide detailed insights into the interplay
between frustration, exchange anisotropy, and DM interactions in triangular
molecular magnets. Beyond their fundamental significance for understanding
magnetism in triangular arrangements, these findings may have practical
implications for the development of molecular-scale magnetic refrigeration
and energy-efficient cooling technologies.
\begin{acknowledgments}
The authors would like to thank CNPq, Capes and FAPEMIG for financial
support. J. Torrico thanks CNPq (163000/2020-4) for partial financial
support. O. R. and S. M. de Souza thanks CNPq and FAPEMIG for partial
financial support. 
\end{acknowledgments}

\end{document}